\documentclass[twocolumn,showpacs,preprintnumbers,amsmath,amssymb]{revtex4}

\usepackage{graphicx}
\usepackage{dcolumn}
\usepackage{bm}

\begin{document}


\title{Appearance of Light Clusters in Post-bounce Evolution of Core-Collapse Supernovae}

\author{Kohsuke Sumiyoshi}
\affiliation{%
Numazu College of Technology, 
Ooka 3600, Numazu, Shizuoka 410-8501, Japan}%

\author{Gerd R\"opke}
\affiliation{
University of Rostock, Institut f\"ur Physik,
Universit\"atsplatz 1, 18051 Rostock, Germany}%

\date{\today}

\begin{abstract}
We explore the abundance of light clusters in core-collapse 
supernovae at post-bounce stage in a quantum statistical approach.
Adopting the profile of a supernova core from detailed numerical 
simulations, we study the distribution of light bound clusters 
up to alpha particles (2 $\leq A \leq$ 4) as well as heavy nuclei ($A > 4$)
in dense matter at finite temperature.
Within the frame of a cluster-mean field approach, the abundances of 
light clusters are evaluated accounting for self-energy, Pauli blocking 
and effects of continuum correlations.
We find that deuterons and tritons, in addition to $^3$He and $^4$He, 
appear abundantly in a wide region from the surface of the proto-neutron 
star to the position of the shock wave.
The appearance of light clusters may modify the neutrino emission 
in the cooling region and the neutrino absorption in the heating region, and, 
thereby, influence the supernova mechanism. 
\end{abstract}

\maketitle

The formation of clusters in nuclear matter is one of pivotal phenomena 
in nuclear structure and heavy ion collisions.
Light clusters such as $^4$He appear also in astrophysical 
environments at certain densities and temperatures.
If they exist abundantly, 
they may change the properties of dense matter 
and contribute to new reactions controling the dynamical 
evolution of stars.

In the explosion mechanism for core-collapse supernovae, 
which has been a long standing problem 
despite continuous research for decades \cite{bet90}, 
the existence of $^4$He and nuclei in the heating region 
has been proposed to be a possible agent 
to help the stalled shock wave 
via the neutrino heating mechanism \cite{hax88,bru91}.
In most of the recent numerical simulations of core-collapse supernovae 
studying the gravitational collapse of massive stars, 
the shock wave produced by a core bounce stalls on the way 
and needs some mechanism to revive the outward propagation of the shock wave.
The heating source through neutrino reactions in dense matter 
behind the stalled shock is one of key issues 
in sophisticated numerical simulations \cite{bur07,jan07}.  
It has been shown that neutrino-$^4$He reactions 
may marginally affect the standing accretion shock instability 
and consequently the supernova shock revival \cite{ohn07}.  

Besides $^4$He, 
the light clusters such as deuterons, tritons and $^3$He 
have not been included in numerical simulations of supernovae so far.  
Those light clusters may become new agents 
for absorption and emission of neutrinos, 
which can carry 
energy away and heat the material 
in the important phase of supernova dynamics.  
In the currently available sets of equations of state (EOS) 
for supernova simulations 
such as the Shen-EOS \cite{she98b} and the Lattimer-Swesty EOS \cite{lat91}, 
dense matter is described as a mixture of neutrons, 
protons, $^4$He, and one representative species of nuclei.  
Since the occurrence of light clusters in addition to $^4$He
is a natural outcome in nuclear statistical equilibrium, 
one should clarify whether they are also present in realistic profiles 
of a supernova core.  
A consistent description of the composition of hot and dense nuclear matter 
is preferable for the rigorous treatment of neutrino transport 
in the post-bounce evolution of core-collapse supernovae.
Progress in the description of clusters 
in dense matter \cite{roe05,roe06} based on a generalized
Beth-Uhlenbeck approach \cite{roe82,sch90}
enables us to evaluate the abundance of deuterons, tritons, 
and helium nuclei altogether within a microscopic approach.  
The mixture of tritons and $^3$He and their neutrino reactions 
have recently been studied under the condition of fixed temperatures 
and proton fractions \cite{oco07}, but without deuterons and heavy nuclei. 

In this paper, we report the appearance of light clusters 
in dense matter of supernova cores adopting the microscopic 
treatment of clusters developed within a quantum statistical approach.  
We aim at the unification of the quasiparticle
approach such as the relativistic mean-field theory \cite{she98b,typ05}, 
which may be adequate at high baryon densities, and the
cluster approach including the 
virial expansion \cite{oco07,roe82,roe83,sch90,hor06}
which may be applicable in the low-density 
limit.

The EOS of nuclear matter $n_i(T,\mu_p,\mu_n)$ giving the total
particle densities of protons and neutrons, $i=p,n$, respectively,
as function of temperature $T$ and chemical potentials $\mu_i$,
is obtained from the spectral function $A(1,\omega)$, 
with $1 = \{\vec k_1, \sigma_1, \tau_1 \}$ denoting momentum,
spin and isospin of a single nucleon. 
In the low-density, low-temperature region, 
a nonrelativistic Hamiltonian approach is possible which is based on 
a nucleon-nucleon interaction $V(12,1'2')$. Well known 
examples are the Bonn and Paris potentials or its separable
representations, reproducing
empirical properties such as scattering phase shifts.
Within this quantum statistical 
approach, the baryon density is calculated as
\begin{equation}
\label{eos}
n_B(T,\mu_p,\mu_n) = \sum_1 \int \frac{d \omega }{2 \pi} \frac{1 }{ {\rm e}^{(\omega-\mu_1)/T}+1} A(1,\omega)\,. 
\end{equation}
Further thermodynamic quantities such as pressure $p(T,\mu_p,\mu_n)$ or 
free energy density $f(T,n_p,n_n)$, being potentials,
are obtained by integrations. 

The spectral function is related to the 
self-energy $\Sigma$,
\begin{equation}
A(1,\omega)=\frac{2 {\rm Im} \Sigma(1,\omega -i0) }{ [\omega - p_1^2/2m_1-{\rm Re} \Sigma(1,\omega)]^2
+[{\rm Im} \Sigma(1,\omega -i0)]^2}\,.\nonumber
\end{equation}
Besides the quasiparticle 
contribution at $E^{\rm qu}(1)=p_1^2/2m_1+{\rm Re} \Sigma(1,E^{\rm qu}(1))$
 it contains also the contribution of bound states and
continuum correlations included in ${\rm Im} \Sigma(1,\omega -i0)$, so that within a generalized 
Beth-Uhlenbeck approach
a virial expansion can be performed in the low-density region.
After the cluster decomposition of the self-energy \cite{roe82,roe83}, the
contributions of the different clusters are 
treated in a systematic way.

The cluster decomposition for the 
self-energy involves the calculation of the $A$-nucleon T matrix
which obeys a corresponding Bethe-Salpeter equation.
Two-particle correlations in nuclear matter can be
accounted for by considering the two-particle Green function in ladder
approximation. The corresponding Bethe-Salpeter
equation contains self-energy and Pauli-blocking terms \cite{sch90}.
The cluster-mean field approximation \cite{roe83} yields a wave equation
which describes also arbitrary clusters in a clustered medium.
For the two-particle system, neglecting correlations
in the surroundings, the following wave equation results
\begin{eqnarray}
\label{wave} 
&&\left[ E^{\rm qu}(1) +E^{\rm qu}(2) -E^{\rm qu}_{2,\lambda}(P) \right]
\psi_{\lambda, P}(12)\\ && +\left[1-f(1)-f(2) \right]\;\sum_{1'2'}
V(12,1'2')\;\psi_{\lambda, P}(1'2') = 0 \,\,.  \nonumber
\end{eqnarray}
The medium Fermi distribution function $f(1)=\{\exp[E^{\rm
qu}(1)/T-\hat \mu_1/T] +1 \}^{-1}$ contains the effective chemical 
potential $\hat \mu_1$ which is determined 
by the total proton or neutron density calculated in quasiparticle approximation, 
$n_i = \sum_1 f(1) \delta_{\tau_1,i}$. It describes the 
occupation of the phase space neglecting any correlations in the medium. 
For the parameter values $T, n_B$ considered below the
account of correlations in the medium gives no significant changes in 
the spectral function \cite{roe83}.

The quasiparticle energies $E^{\rm qu}(1)$ are self-consistently calculated 
if we evaluate the self-energy or the corresponding T matrix. Using the cluster
decomposition and restricting only to the two-particle contribution \cite{sch90}, the results
coincide with the Brueckner K matrix including hole-hole contributions.
The quasiparticle energies can be parametrized and directly related to
empirical data on finite nuclei. An appropriate parametrization can be found
on the basis of relativistic mean-field theories \cite{she98b,typ05} which
also involves the results of relativistic Brueckner-Hartree Fock 
calculations \cite{mar07}. We will use the TM1 parametrization here.

From the solution of this in-medium two-particle Schr\"odinger
equation (\ref{wave}) the scattering states and possibly
the bound states are obtained. Due to the
self-energy shifts and the Pauli blocking, the binding energy of the
deuteron $E_d(P;T,\mu_p, \mu_n)$ as well as the scattering phase
shifts $\delta_{\lambda}(E,P;T,\mu_p, \mu_n)$ in the respective isospin singlet or
triplet channel $\lambda$, will depend on the temperature
and the chemical potentials \cite{sch90,ste95}. 

In-medium Schr\"odinger equations similar to Eq.~(\ref{wave}) containing
quasiparticle shifts and Pauli blocking terms are derived for higher 
clusters with mass number $A$ and charge $Z$ \cite{roe83}.
The shift of the bound state energies $E^{\rm qu}_{A,Z}(P;T,n_i)
= E_{A,Z}+\Delta_{A,Z}^{\rm SE}+\Delta_{A,Z}^{\rm Pauli}
+\Delta_{A,Z}^{\rm Coul}$, containing in addition to the single-particle
self-energy shift $\Delta_{A,Z}^{\rm SE}$ and the Pauli blocking term 
$\Delta_{A,Z}^{\rm Pauli}$ also the Coulomb shift 
$ \Delta_{A,Z}^{\rm Coul} $, can be calculated within 
perturbation theory. Besides the quasiparticle shift at zero momentum
$\Delta_{A,Z}^{\rm SE,0} = Z E^{\rm qu}_{p}(0)+(A-Z) E^{\rm qu}_{n}(0)$
which can be included into the chemical potential, there is the contribution due to the
effective mass $m^*$. Assuming for light clusters $2\le A \le 4$ a Gaussian wave 
function with the nucleonic rms radii $\langle r^2\rangle_{A,Z}$, the
self-energy shift results as $\Delta_{A,Z}^{\rm SE}= \Delta_{A,Z}^{\rm SE,0} 
+ 3(A-1)\hbar^2 b^2_{A,Z}(m-m^*)/(8m^2)$ after introducing 
Jacobian coordinates and separating the c.o.m. motion,
where $b^2_{A,Z} = 3 (A-1)/(A \langle r^2 \rangle_{A,Z})$.
With Gaussian wave functions we obtain the following estimate for the
Pauli blocking shift
\begin{equation}
\label{delpauli}
\Delta_{A,Z}^{\rm Pauli}=\int \frac{q^2 dq {\rm e}^{-\frac{2Aq^2}{b^2(A-1)}} 
[|E|+\frac{\hbar^2}{2m}(\frac{A}{A-1}q^2+\frac{3(A-2)}{4}b^2)]}{({\rm e}^{[\frac{\hbar^2}{2m}
(\frac{P}{A}+q)^2-\hat \mu_i]/T}+1) \left[\frac{A-1}{A} \right]^{3/2} \frac{\sqrt{\pi}}{8} b^3}\,\,.
\end{equation}

Similar expressions \cite{roe84} can be given for the weakly bound clusters with 
$5 \le A \le 11$. The Coulomb shift $\Delta_{A,Z}^{\rm Coul}$ 
is calculated in Wigner-Seitz approximation, 
see also \cite{lat91,roe84}. Within the 
parameter values considered below, the influence of the Coulomb corrections 
on the composition is small.

For heavier clusters with 
mass numbers $A \ge 12$, the self-energy and Pauli-blocking shifts become less 
important and will be neglected here. The heavier clusters repel
the nuclear matter so that the mean-field effects are mostly produced by the other 
nucleons within the cluster and are contained in the cluster binding
energy. For a more detailed consideration see \cite{roe84}. 

Using the cluster decomposition of the self-energy, from Eq. (\ref{eos}) a
generalized Beth-Uhlenbeck formula can be derived which is formally
similar to the virial expansion of the EOS
\begin{equation}
\label{n12}
n_B(T,\mu_p,\mu_n)= \sum_{A=1,Z}^\infty A\,\, n_{A,Z}(T,\mu_p,\mu_n)\,.
\end{equation}
Introducing the Fermi or Bose distribution function\\
$ f_{A,Z}(E) =\left[ {\rm e}^{(E-Z \mu_p-(A-Z) \mu_n)/T}-(-1)^A\right]^{-1}$,
the EOS (\ref{n12}) contains the single-particle contribution
$n_{1,Z}= 2 \sum_P f_{1,Z}(E^{\rm qu}(P))$,
the two-particle contributions $n_{2,Z} = n_{2,Z}^{\rm bound} + n_{2,Z}^{\rm scat}$
(consisting of the contribution of medium-modified deuterons
and scattering states in the isospin singlet and
triplet channel, see \cite{ste95,sch90}), as well as the contribution 
of all higher clusters.

The bound state contribution of the $A$-nucleon cluster to the EOS (\ref{n12})
has the form \cite{roe84,sch90,hor06} $ n_{A,Z}
= \sum_{P,\lambda}\left[ f_{A,Z}(E^{\rm qu}_{A,Z,\lambda}(P))-f_{A,Z}(E^{\rm cont}_{A,Z}(P)) \right]
$
where $\lambda$ denotes the internal degrees of freedom such as spin and excitation states.
The summation over the c.o.m. momentum $P$ runs only over the region where the in-medium
binding energy $E^{\rm qu}_{A,Z,\lambda}(P)$ is lower than the edge of the continuum 
$E^{\rm cont}_{A,Z}(P) = A\,\, E^{\rm qu}(P/A)$.
The summation 
over all excited states of a large cluster can be replaced by an intergal over the excitation energy 
using the density of states, see \cite{roe84,lat91}.

Of importance is the contribution of continuum correlations which can be represented 
by the scattering phase shifts. Extended investigations have been performed with
respect to the contribution of two-nucleon phase shifts \cite{sch90}. They reduce the 
contribution of the bound states. Partial integration, using the Levinson theorem,
gives the second contribution $ f_{A,Z}(E^{\rm cont}_{A,Z}(P)) $. 
Thus, there is no jump in the EOS when a bound state merges with the continuum 
if, e.g., the baryon density is increased.
The remaining correlations in the continuum are compensated to a large extent 
by the contribution of interaction to the 
quasiparticle shift 
and will be neglected. Explicitly this can be shown for the contribution of
the two-nucleon singlet and triplet scattering phase shift contributions.

In order to explore the fraction of nucleons bound in light clusters, 
we utilize a snapshot of a supernova core at post bounce stage, 
which is important for setting the revival condition
of a shock wave and the resulting explosion.  
We use the profile obtained by the detailed 
numerical simulation \cite{sum05} 
of the gravitational collapse and core bounce 
for a 15M$_{\odot}$ pre-supernova star 
adopting the Shen EOS \cite{she98b}.  

\begin{figure}
\includegraphics[width=8cm]{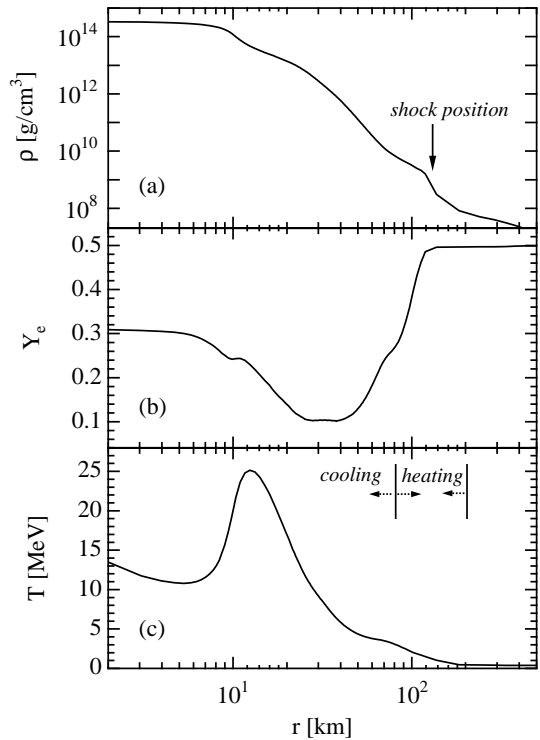}
\caption{\label{fig:profile}
Density (a), electron fraction (b) and 
temperature (c) profile of the supernova core at 150 ms after 
core bounce as a function of the radius.
The position of shock wave and 
the boundaries of heating and cooling regions are 
indicated in the top and bottom panels.}
\end{figure}

Figure \ref{fig:profile} displays the profiles of 
density ($\rho$), electron fraction ($Y_e$) and temperature ($T$) 
at 150 ms after the core bounce. 
The shock wave is stalled at $\sim$130 km. 
The proto-neutron star is just born at the center and matter is 
falling down onto it through the shock wave.  
The neutrino sphere, where the neutrinos are emitted, 
is located at 20--80 km depending on energies and species.  
The electron fraction is $\sim$0.5 in the iron nuclei region at $r \ge$100 km 
and becomes smaller (neutron rich) for dissociated nucleons inside 
due to the balance under beta equilibrium including neutrinos.
Note that the electron fraction is the same as the total proton fraction
due to charge neutrality.  

At the shock wave, the iron nuclei are dissociated into $^4$He, 
and, subsequently, nucleons at high temperatures due to the shock heating.  
We remark that the nuclear statistical equilibrium is maintained 
during the dynamics 
because the temperature and density are high at the region of current 
interest and, therefore, the compositional change proceeds fast enough
via charged particle reactions.  
During this process of dissociation, light clusters may appear 
in addition to $^4$He and can modify the cooling 
and heating rates through weak reactions with leptons.  
Conventionally, only nucleons are considered for the cooling 
process via neutrino emission at the surface of proto-neutron 
star (cooling region) 
and the heating process via neutrino 
absorption behind the shock wave (heating region).

\begin{figure}
\includegraphics[width=9cm]{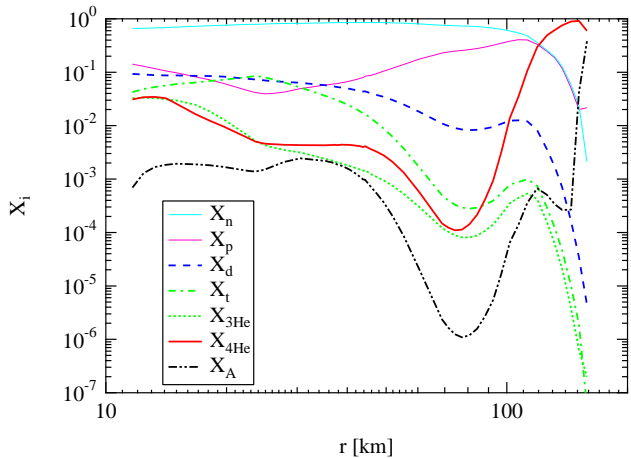}
\caption{\label{fig:fraction} (Color online) 
Mass fraction $X_{i}$ of light clusters 
 as function of the radius for the post-bounce supernova core 
 shown in Fig. 1.}
\end{figure}

We solved the EOS (\ref{n12}) for given profiles of $T, \rho, Y_e$ as shown in 
figure \ref{fig:profile}. The masses of nuclei are
taken from Audi and Wapstra \cite{aud93}.
The chemical potentials $\mu_p, \mu_n$ are determined in a self-consistent way.
The abundances of the different clusters have been calculated, and 
fractions $X_{A,Z}=A\,n_{A,Z}/n_B$ of nucleons bound in 
deuterons, tritons, $^3$He, $^4$He as well as the fraction $X_A$ 
of nucleons bound in nuclei with $A \ge 5$ are shown in figure \ref{fig:fraction}.

The contributions of free protons and free neutrons to the baryon density
follows nearly the behavior given by Shen et al. \cite{she98b,sum05}.
In the inner region ($r < 30$ km) where the density and temperature become high, 
dominant clusters are deuterons and tritons because of the high neutron fraction. 
Here the main contribution to $X_A$ is from $A=5$.
In the outer region ($r > 100$ km), 
$^4$He and heavy nuclei become abundant.  
In the region between, the deuteron fraction is relatively high 
and the fraction of $^4$He rises to about 90 \% at $r \sim 150$ km, 
before it goes down 
when the heavier clusters, $A \ge 5$, are formed. The wiggle of $X_A$ near $r \sim 150$ km is 
caused by the weak binding energy of nuclei with $5 \le A \le 11$ which are of relevance there.
At larger radii, the strongly bound nuclei of the iron region will dominate.

The appearance of light clusters in supernova cores at post-bounce stage 
indicates that there are additional weak reactions involving them 
which will have 
influence on the supernova dynamics.  
The neutrino interactions on light nuclei, which appear 
inside the proto-neutron star at $\sim$10 km, will contribute to 
the opacity.
This may lead to a modification of the diffusion coefficient for neutrinos, 
the time scale of neutrino diffusion and the location of neutrino sphere, 
where most of the neutrinos start streaming freely.  
The light nuclei at $\sim$60 km in the cooling region 
are possible sources of neutrino emission 
through electron and positron captures.  
They will modify the neutrino luminosities and spectra 
because the emission rates per nucleon for light nuclei 
may be effectively smaller 
than the large emission rates for nucleons.
Medium effects on the neutrino emissivity have been 
considered in \cite{bla95} 
using the many-particle approach described above.  
In fact, the abundance of nucleons is reduced by about 10\% 
due to the appearance of light clusters and 
the corresponding neutrino luminosities may be lowered.
We expect that 
these modifications of the neutrino flux and energy change 
the neutrino heating rates.  

The light clusters in the heating region behind the shock wave 
become the target compositions for neutrino absorption 
and will modify the heating rate to assist the revival of shock wave.  
The dominance of $^4$He at $\sim$150 km in the heating region 
may contribute to the heating rate via neutrino 
absorption as discussed by Haxton et al. \cite{hax88,bru91}.
It is interesting to see that 
the fraction of $^4$He in the current approach 
is larger than the value in the Shen EOS \cite{sum05}
where the low binding energies for the clusters 
with 5 $\le$ A $\le$ 11 are not accounted for.  
Deuterons become more abundant than $^4$He at $r < 100$ km 
and dominate over other nuclear species in the heating region.
The neutrino absorption cross section on deuteron is 
much larger than that for $^4$He \cite{yin89,nak01,nak02}, 
therefore, deuterons may become more effective targets 
for the heating than $^4$He.  
Neutrino heating processes on triton and $^3$He 
should be studied whether they may contribute as well \cite{oco07}.  
When nucleons appear together with light clusters, 
the heating via the neutrino absorption on nucleon 
is dominant since the absorption cross section on nucleon 
is largest among others.  
In this case, the appearance of light clusters reduces 
the total heating rate.  
These processes with light clusters 
will influence the neutrino-heating mechanism 
in the revival of the shock wave for a successful explosion.

We have adopted the spherically symmetric snapshot 
at 150 msec after bounce as an exemplary case.  
Other time sequences and multi-dimensional aspects 
will be studied in our subsequent paper 
to examine whether light clusters appear generally in a significant amount.
It is known that nuclei appear more abundantly behind the shock wave 
in the hydrodynamical instability and the neutrino-$^4$He reactions 
may contribute to the heating \cite{ohn07}.  
It would be interesting to study the fractions of light 
clusters in profiles with wider ranges of density 
and temperature in multi-dimensional simulations.
When the reaction time scale becomes 
too long to achieve the equilibrium in fast flows, 
one needs to treat further 
the dynamical formation or destruction of light clusters.  

Finding the light clusters in supernova cores motivates 
detailed studies on weak reactions regarding light clusters.
Their reactions must be taken into account, in principle,
in any neutrino transfer simulation of core-collapse supernovae 
and proto-neutron star cooling.  
It would be preferable to construct the supernova EOS table 
taking into account the formation of light clusters. The approach 
given here merges a low-density virial expansion with a quasiparticle 
approach at higher densities.
Since the appearance of light clusters is a natural outcome 
in dense matter, their existence has to be considered 
in critical studies not only of the supernova mechanism, but also
other problems in nuclear physics.

The authors express their sincere gratitude for the hospitality 
of Max Planck Institut f\"ur Astrophysik, where the new idea 
to write this paper comes out during their research stay.  
They are grateful especially to W. Hillebrandt and H.-T. Janka 
for fruitful discussions and local supports.  
K. S is grateful to K. Kotake, S. Yamada, H. Suzuki, H. Shen 
and H. Toki for the collaborations on the equation of state 
and the supernova simulations.  
G.R. thanks D. Blaschke, J. Natowitz, P. Schuck, 
S. Shlomo, S. Typel and H. Wolter for discussions. 
The numerical simulations were performed 
at NAOJ (iks13a, uks06a), JAEA and YITP.
This work is partially supported by the Grants-in-Aid for the 
Scientific Research (18540291, 18540295, 19540252) 
of the MEXT of Japan.

\end{document}